\documentclass[aps,elsarticle,showpacs,superscriptaddress]{revtex4}  % for double-spaced preprint
\usepackage{graphicx}  % needed for figures
\usepackage{dcolumn}   % needed for some tables
\usepackage{bm}        % for math
\usepackage{amsmath}   % for math
\usepackage{subfigure} % for subfigures
\usepackage{hyperref}% for urls
\preprint{
\hbox to \hsize{
\hfill$\vcenter{\hbox{\bf MADPH-11-1567}}$}
}
\begin{document}

\title{Tevatron Asymmetry of Tops in a W', Z' Model}
\author{Vernon~Barger}
\email{barger@pheno.physics.wisc.edu}
\affiliation{Department of Physics, University of Wisconsin, 1150 University
Avenue, Madison, Wisconsin 53706 USA}

\author {Wai-Yee Keung}
\email{keung@fnal.gov}
\affiliation{Department of Physics, University of Illinois, Chicago, IL 60607-7059 USA}
\affiliation{Department of Physics, Brookhaven National Laboratory, Upton, NY 11973, USA}

\author{Chiu-Tien Yu}
\email{cyu27@wisc.edu}
\affiliation{Department of Physics, University of Wisconsin, 1150 University Avenue, Madison, Wisconsin 53706 USA}

\date{February 1, 2011;~updated February 24, 2011}

\begin{abstract}
Recent results from the CDF Collaboration indicate that the top-pair forward-backward asymmetry is largest in regions of high rapidity difference $|\Delta y|$ and invariant mass $M_{t\bar t}.$ We show that experimental observations can be explained by our previously proposed Asymmetric Left-Right Model (ALRM).  The gauge symmetry $U'(1)\times SU'(2)\times SU(2)$ is broken by a triplet Higgs in the primed sector.  The $W'$ boson has a $(t,d)$ right-handed coupling and the $Z'$ boson has diagonal fermion couplings.  We determine the model parameters to be $M_{W'} = 700$ GeV, $M_{Z'}=1$ TeV, and charged current coupling $g'_2=3$.  The $W'$ and $Z'$ total decay widths are of $\mathcal{O}(100$ GeV).  The signals from $Z'\to t\bar t$ and $W'\to tb$ at the LHC will test the model.\end{abstract}

\maketitle
\section{Introduction}
The top-pair forward-backward asymmetry at the Tevatron collider by the CDF collaboration \cite{CDFAfbOld} has generated much interest \cite{miscref} as a possible harbinger of new physics. At lowest order in the SM, the top-pair production is symmetric under charge conjugation. A small forward-backward asymmetry of $A_{FB}=0.06\pm0.01$ \cite{Afb_SM} in the $t\bar t$ rest-frame arises through the interference of NLO QCD processes that differ under charge conjugation. Recent results from the CDF Collaboration \cite{CDFAfbNew} using a data set of 5.3 fb$^{-1}$ show that the forward-backward asymmetry in top pair production $A_{FB}$ still deviates from SM predictions and, furthermore, is mass-dependent. In particular, the asymmetry is most prominent in regions of high rapidity difference $|\Delta y|>1$ and large invariant mass $M_{t\bar t}>450$ GeV, where there are $2\sigma$ and $3\sigma$ deviations,  respectively, from NLO predictions. Such a distribution is a generic feature in the $t$-channel exchange of a particle  \cite{KCnew} \cite{Barger:2010mw} \cite{Shu}\cite{Jung}. In our previous work \cite{Barger:2010mw}, we proposed an Asymmetric Left-Right Model (ALRM) based on the gauge symmetry $U'(1)\times SU'(2)\times SU(2)$. This model has electroweak $W'$ and $Z'$ bosons in addition to the $W$ and $Z$ of the SM. In that work, we did not assume a specific Higgs mechanism, but treated the $W'$ and $Z'$ masses to be independent parameters. In the present paper, we assume that the primed gauge group is broken by the neutral member of a triplet Higgs sector by a vacuum expectation value (vev) $v'$. The SM electroweak gauge group is broken as usual by the vev $v$ of the neutral member of the SM Higgs doublet. In the ALRM, the $W'$ boson has a $(t,d)$ right-handed coupling with coupling $g'_2$ and the $Z'$ boson has diagonal fermion couplings. We determine the model parameters from the CDF data to be $M_{W'}=700$ GeV, $M_{Z'}=1$ TeV, and $g'_2=3$. The $W'$ and $Z'$ total decay widths are then of $\mathcal{O}(100\mathrm{~GeV}).$

We begin with an overview of the ALRM and then discuss the symmetry breaking in Section II. In Section III, we calculate the forward-backward asymmetry $A_{FB}$ as functions of $|\Delta y|$ and $M_{t\bar t}$ and compare our results with the CDF data. In Section IV, we evaluate signals of the ALRM at the LHC.  The signals at the LHC from  $Z', W'$ production and their $Z'\to t\bar t$, $W'\to tb$ decays can definitively test this model interpretation of the CDF top asymmetry data.  We end with our conclusions in Section V.
 
\section{Asymmetric Left-Right Model}
The ALRM begins with the gauge group $U'(1)\times SU'(2)\times SU(2)$.
The unprimed $SU(2)$ acts on the usual SM left-handed quark doublets.
The primed $SU'(2)$ applies to the right-handed
doublet $(t,d)_R^T$ in an unconventional grouping. 

The gauge symmetries are broken
sequentially, starting with $U'(1)\times SU'(2)
\to U_Y$ to obtain the SM hypercharge ${Y\over2}=T'_3+{Y'\over2}$, and
then $U_Y(1)\times SU(2)
\to U_{EM}$ to obtain  $Q=T_3+{Y\over2}$. By using this sequential approximation to the
breaking, we preserve the SM interaction.

After symmetry breaking, there are massive $Z'$ and $W'^\pm$ bosons in
addition to the usual weak bosons. The ${W'^\pm}$ have a $Z'W'W'$ tri-gauge boson coupling given by $g_2'^2/\sqrt{g'^2+g_1'^2}$, which is of order ${\cal O}(1)$.
In order to preserve unitarity, the SM $Z$ also appears in
the vertex $ZW'^+W'^-$ with coupling   $-e\tan\theta_W$. Additional massive fermions will be needed for anomaly cancellation.

Our results are derived by making two successive rotations of gauge boson states. First, we make the rotation
\begin{equation}
 B = (g'_2B' + g_1' W'^3)/\sqrt{{g_1'}^2+g_2'^2} \ ,\\ \newline
  Z'= (-g_1' B' + g'_2 W'^3)/\sqrt{{g_1'}^2+g_2'^2}  \ .
\end{equation}
Then the couplings to the generators become
\begin{equation}
 g_1' { Y'\over2}  B' +  g'_2 T'_3 W'_3
 =\left({g_1' g'_2\over \sqrt{{g_1'}^2+g_2'^2}}\right) {Y\over2} B
  +  \sqrt{g_1'^2+g_2'^2} \left( T'_3 -{g_1'^2{Y\over2}\over g_1'^2+g_2'^2}
   \right) Z' \ .
\end{equation}
Subsequently, we perform the usual SM rotation from the basis of $B,W^3$ to
the basis of $A,Z$. To simplify the expressions, we denote
%%%
$ g'=\sqrt{g_2'^2+g_1'^2}$. The SM hypercharge coupling is
$g_1=g_1'g_2'/g'=e/c_W$, and the SM $SU(2)$ coupling is $g_2=e/s_W$.
It is also useful to note that
{$g_1'=(c_W^2/ e^{2}-g_2'^{-2})^{-{1\over2}} $}.

\subsection{$W', Z'$ Mass Relation}
We assume that the Higgs mechanism is due to the condensate of 
a primed Higgs triplet with $T'=1$,
which is denoted as $\phi'$ with its vev as $(0,0,v'/\sqrt{2})^T$. Using a Higgs triplet allows for a larger mass gap between the $Z'$ and $W'$ bosons than a Higgs doublet.
Therefore,
\begin{equation}\hbox{$1\over2$}  M_{Z'}^2 = 
\langle \phi'^\dagger  (g'T_3')^2 \phi' \rangle \ ,\quad
M_{Z'}=g'v' \label{mzp}\end{equation}
%%%
and
\begin{equation} M_{W'}^2=\langle \phi'^\dagger T'_- (g_2/\sqrt{2})^2 T'_+ \phi'\rangle 
\ ,\quad M_{W'}=\hbox{$1\over\sqrt{2}$}g_2'v' \ \label{mwp}.\end{equation}
The two-stage approximation is justified by  $M_{Z'}, M_{W'} \gg M_Z, M_W$. Thus, \begin{equation} {M_{W'}\over M_{Z'}}= {g_2'\over {\sqrt{2}g'}}=\sqrt{\frac{g'^2_2-e^2/c_W^2}{2g'^2_2}}\label{mwpmzprelationTriplet}\end{equation}
 
\begin{figure}[htbp]
\subfigure[~$M_{W'}$ vs. $g'_2$ for $M_{Z'}=800,~900,~1000$ GeV.]{
\includegraphics[scale=1]{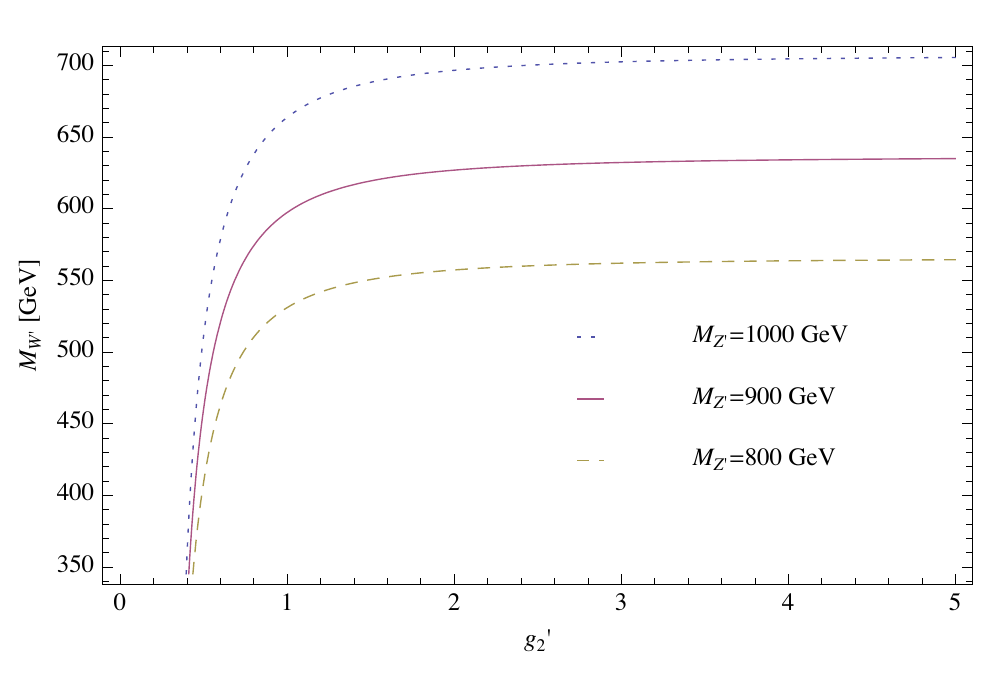}
}
\subfigure[~$M_{W'}/M_{Z'}$ vs. $g'_2$]{
\includegraphics[scale=0.95]{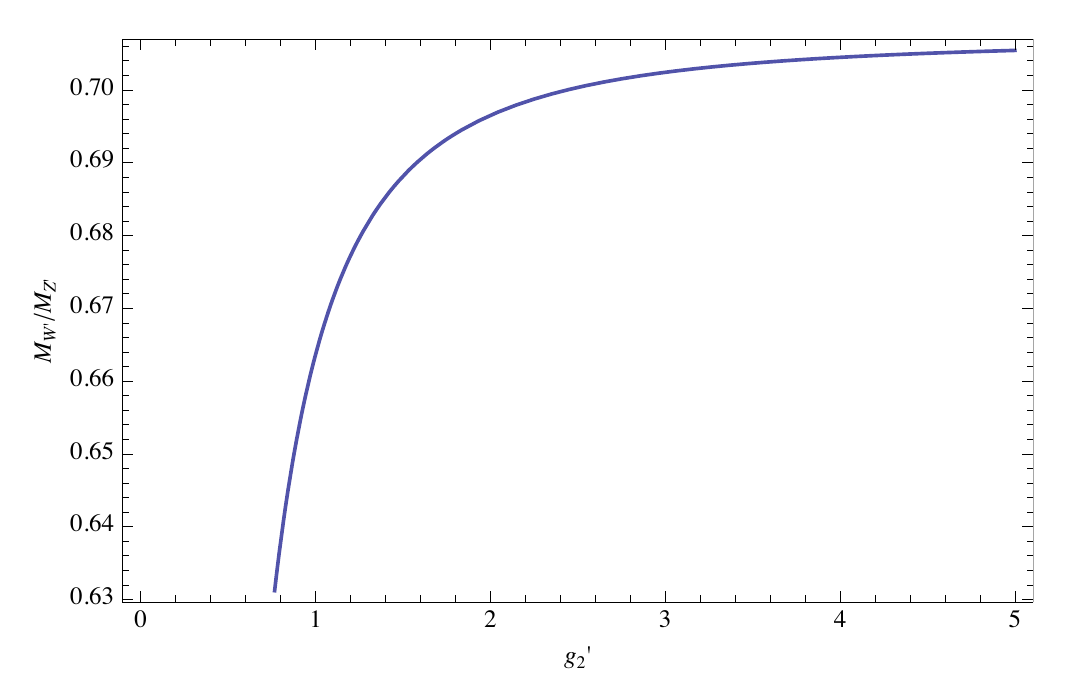}
}
\caption{Dependence of $M_{W'}$ on $g'_2$ and $M_{Z'}$. The relationship is given by Eq. \ref{mwpmzprelationTriplet}}
\label{fig:mass_relation_plot3}
\end{figure}

%In order to satisfy constraints from $Z'$-searches \cite{CDFzprime}, we require that $M_{Z'}\gtrsim 800$ GeV \footnote{The $M_{Z'}$ bound depends on $g'_2$. A larger $g'_2$ decreases the lower bound on $M_{Z'}$. For example, for $g'_2=1$, $M_{Z'}\gtrsim800$ GeV and for $g'_2=3.0$, $M_{Z'}\gtrsim 500$ GeV. We have chosen a conservative value of $M_{Z'}\sim 1$ TeV.}. 
A search by CDF in the muon-pair channel established a lower bound of $M_{Z'}>1051$ GeV for a $Z'$ with SM couplings  \cite{CDFzprime}. Dijet searches have placed exclusion limits of $320<M_{Z'}< 740$ GeV, again with SM coupling of the $Z'$ \cite{CDFzprimedijet}. For our $Z'$, with $g'_2=3$, the dimuon limit is $M_{Z'}\gtrsim 500$ GeV while the dijets do not limit $M_{Z'}$. We plot the relationship between $M_{W'}$, $M_{Z'}$, and $g'_2$ in Fig.\ref{fig:mass_relation_plot3}. Numerically, we use $e^2=4\pi\alpha(M_Z)=0.30$ and $c_W^2=1-0.23=0.77$. After accounting for constraints from experimental data, we use $M_{W'}=700$ GeV, $M_{Z'}=1$ TeV and $g'_2=3$ as our benchmark point. Note that for $g'_2 > 1$, the $M_{W'}/M_{Z'}$ ratio is relatively insensitive to the value of $g'_2$ and it is approximately given by the large $g'_2$ limit of $M_{W'} = M_{Z'}/\sqrt{2}$.

\subsection{$W'$ and $Z'$ decays}
For $M_{W'}=700$ GeV and $g'_2=3$, the width for $W'$ decay is
%\footnote{The large width of the $W'$ makes it more difficult to distinguish from SM backgrounds. Since the invariant mass distributions of the decay products depends on the total width, the broader distributions make new physics processes more difficult to distinguish from SM backgrounds.}      
\begin{equation}
\Gamma (W' \rightarrow t\bar d) = \frac{g_2'^2 }{(16 \pi)}M_{W'}\left(1-\frac{3 m_t^2}{2 M_{W'}^2}+ \frac{m_t^6}{2 M_{W'}^6}\right ) = 114 \mathrm{~GeV}.                
\end{equation}

For $M_{Z'}=1$ TeV and $g'_1=0.35$, the partial widths for the $Z'$ decays are
\begin{eqnarray}                           
\Gamma (Z'\rightarrow d \bar d) &=& \left[
\left(         {g_1'^2\over 6g'}                      \right)^2+
\left(         {g_1'^2\over 3g'}   -{g'\over2}   \right)^2  
\right]   {M_{Z'}\over 8\pi} = 89\mathrm{~GeV}\\
%\Gamma (Z'\rightarrow d \bar d) = \frac{1}{3}{\left(\frac{g_1'^2}{6g'}\right)^2 + \left[\left(\frac{g_1'^2}{3g'}\right)^2 + 0.5g'\right]^2}\frac{M_{Z'}}{(24\pi)} =  0.7 \mathrm{~GeV }               
\Gamma(Z'\to t\bar t)&=&\frac{M_{Z'}}{64\pi}\left[\left(g'-\frac{5g_1'^2}{3g'}\right)^2\left(1+\frac{2m_t^2}{M_{Z'}^2}\right)+\left(g'-\frac{g_1'^2}{g'}\right)^2\left(1-\frac{4m_t^2}{M_{Z'}^2}\right)\right]\\
&\times&\sqrt{1-\frac{4m_t^2}{M_{Z'}^2}}=79 \mathrm{~GeV}\nonumber\\
\Gamma (Z'\rightarrow l \bar l)& =& \left(\frac{3g_1'^2}{2g'}\right)^2 \frac{M_{Z'}}{8\pi} =  0.1\mathrm{~GeV }\\     
\Gamma (Z'\rightarrow u \bar u)& =& \left(\frac{5g_1'^2}{6g'}\right)^2 \frac{M_{Z'}}{8\pi} =  0.04\mathrm{~GeV }\\
\Gamma (Z'\rightarrow \nu \bar \nu)& =& \left(\frac{g_1'^2}{2g'}\right)^2 \frac{M_{Z'}}{8\pi} =  0.01\mathrm{~GeV }                                     
\end{eqnarray}
The total $Z'$ width is $\Gamma_{Z'}=168$ GeV. The variation of the $Z'$ width with $\hat s$ will be approximated in our calculations of $Z'$ production by the prescription $M_{Z'}^2\to\hat s$, where $\hat s$ is the subprocess CM energy. The branching ratios to a top-pair and leptons are
\begin{eqnarray}
\mathcal{BR}(Z'\to t\bar t)&=&0.5\\
\mathcal{BR}(Z'\to l\bar l)&=&6.0\times10^{-4}
\end{eqnarray}
where $l$ is the sum of $e,\mu,$ and $\tau$ modes in the width. The small leptonic width of the $Z'$ along with a broad total width makes its detection in the Drell-Yan lepton channel difficult. 

%%%
\section{Results}
The asymmetry $A_{FB}$ in the $p\bar p$ center-of-mass frame is defined as
\begin{equation}
A_{FB}=\frac{N(\Delta y>0)-N(\Delta y<0)}{N(\Delta y>0)+N(\Delta y<0)}
\end{equation}
where $\Delta y=y_t-y_{\bar t}$ is the difference in rapidities of the top and anti-top quark. We implemented the ALRM in MadGraph/MadEvent 4.4.44 \cite{Alwall:MG/ME}, using CTEQ6.6M parton distribution functions \cite{Nadolsky} with factorization and renormalization scales $\mu_F=\mu_R=m_t$ \cite{Melnikov}. We took $m_t=173.1$ GeV \cite{PDG:Amsler}\cite{Tevatron:topmass} and applied a uniform SM K-factor of $K=1.3$ to approximate the higher order QCD corrections for (NNLO$_{\rm approx.}$)/(LO) as shown in \cite{Kfactor}.

\begin{figure}[htbp]
\begin{center}
\includegraphics[scale=0.5]{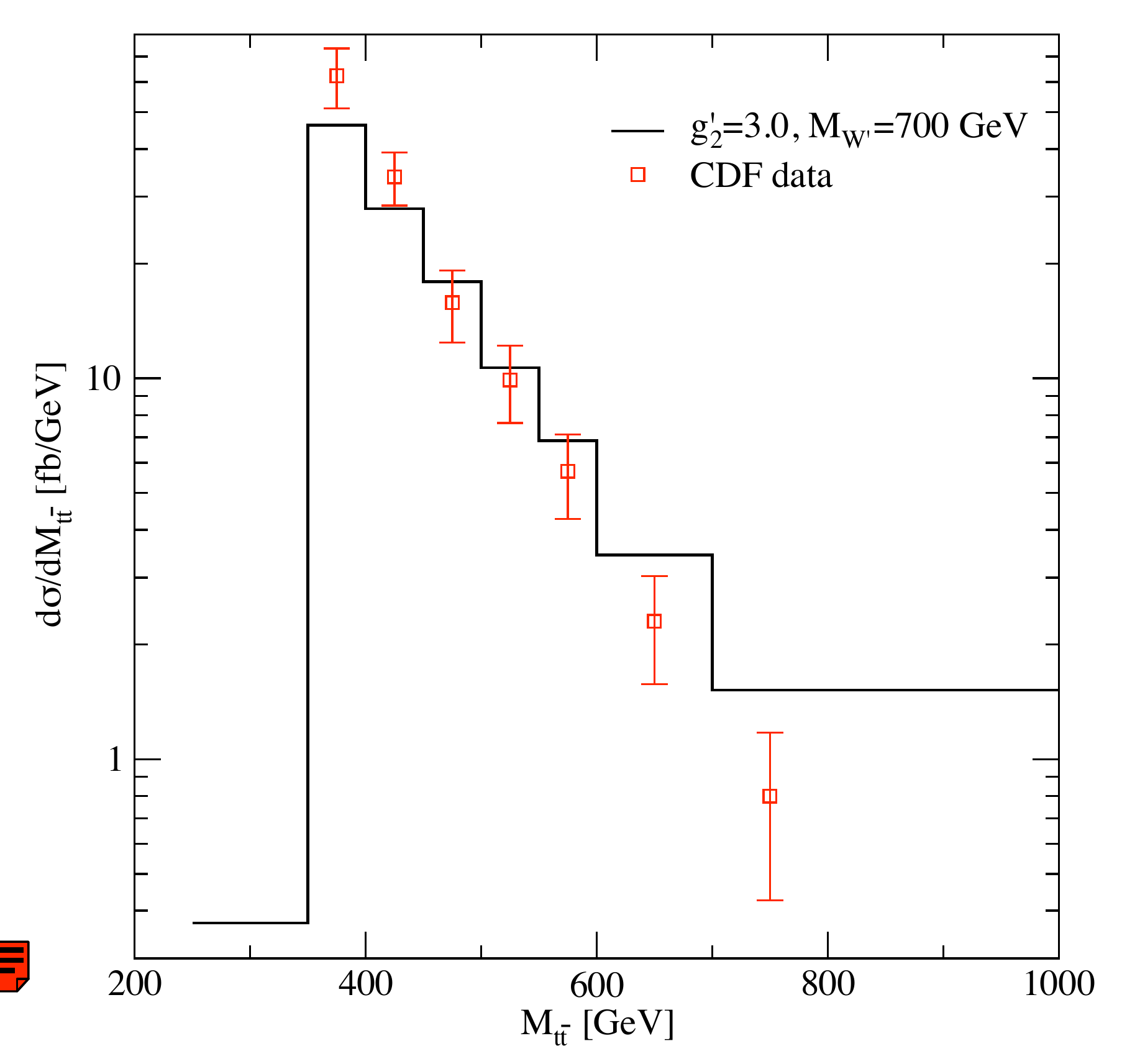}
\caption{$d\sigma/dM_{t\bar t}$ distribution of CDF data(points with uncertainties) vs. ALRM  (solid histograms) for the model benchmark point of $g'_2=3.0,~M_{W'}=700$ GeV, and $M_{Z'}=1$ TeV.}
\label{fig:mtt}
\end{center}
\end{figure}

\begin{figure}[htbp]
\subfigure[]{
\includegraphics[scale=0.4]{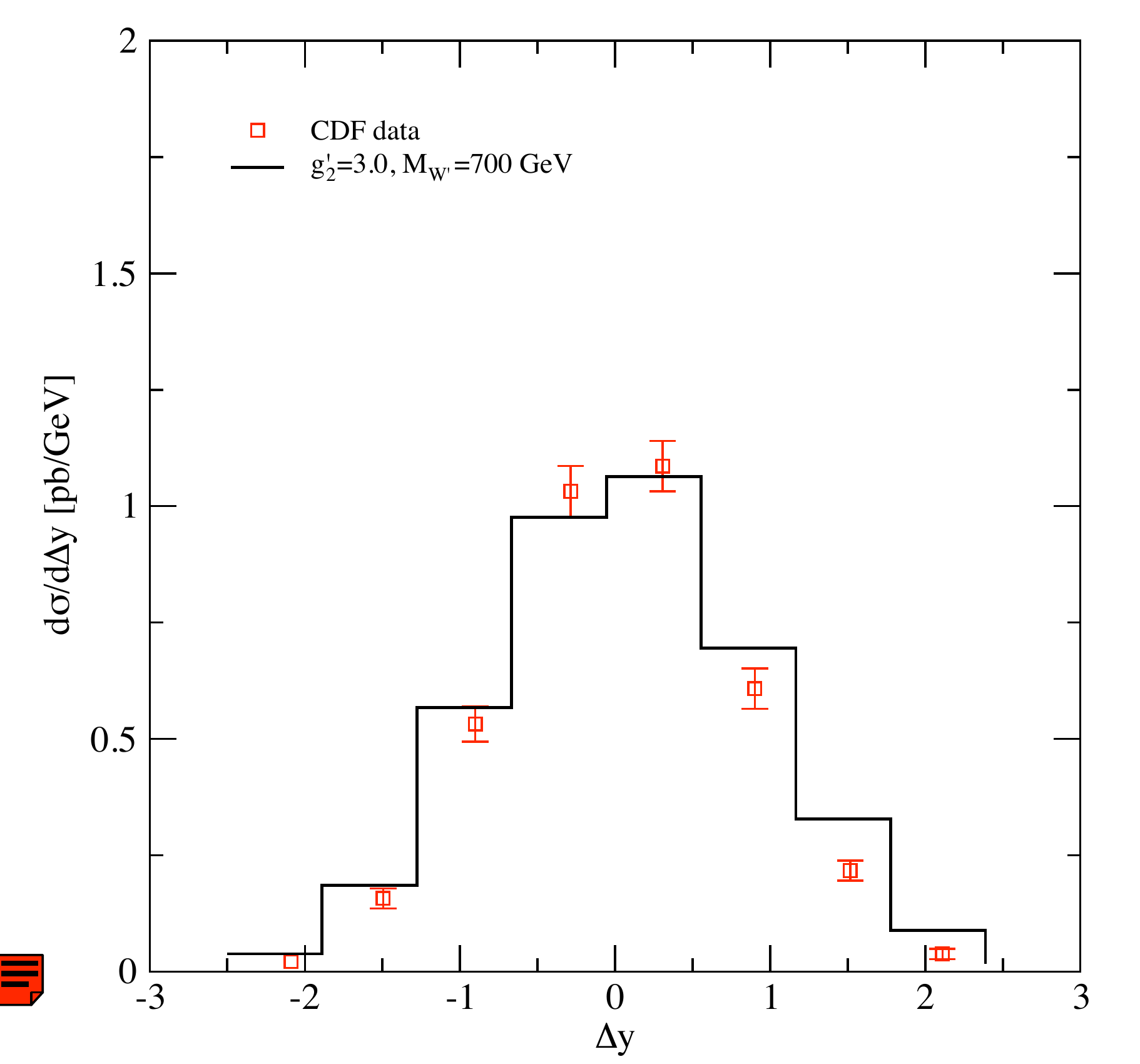}
\label{fig:deltay}
}
%\end{figure}
%\begin{figure}[htbp]
%\caption{Parton level asymmetries at small and large $\Delta y$ for the CDF data and ALRM prediction for $g'_2=3.0,~M_{W'}=700$ GeV, and $M_{Z'}=996.6$ GeV.}{\includegraphics[scale=0.5]{deltay_afb_best1}
%\label{fig:rapidity_afb}
%}
%\end{figure}
\subfigure[]{
\includegraphics[scale=0.4]{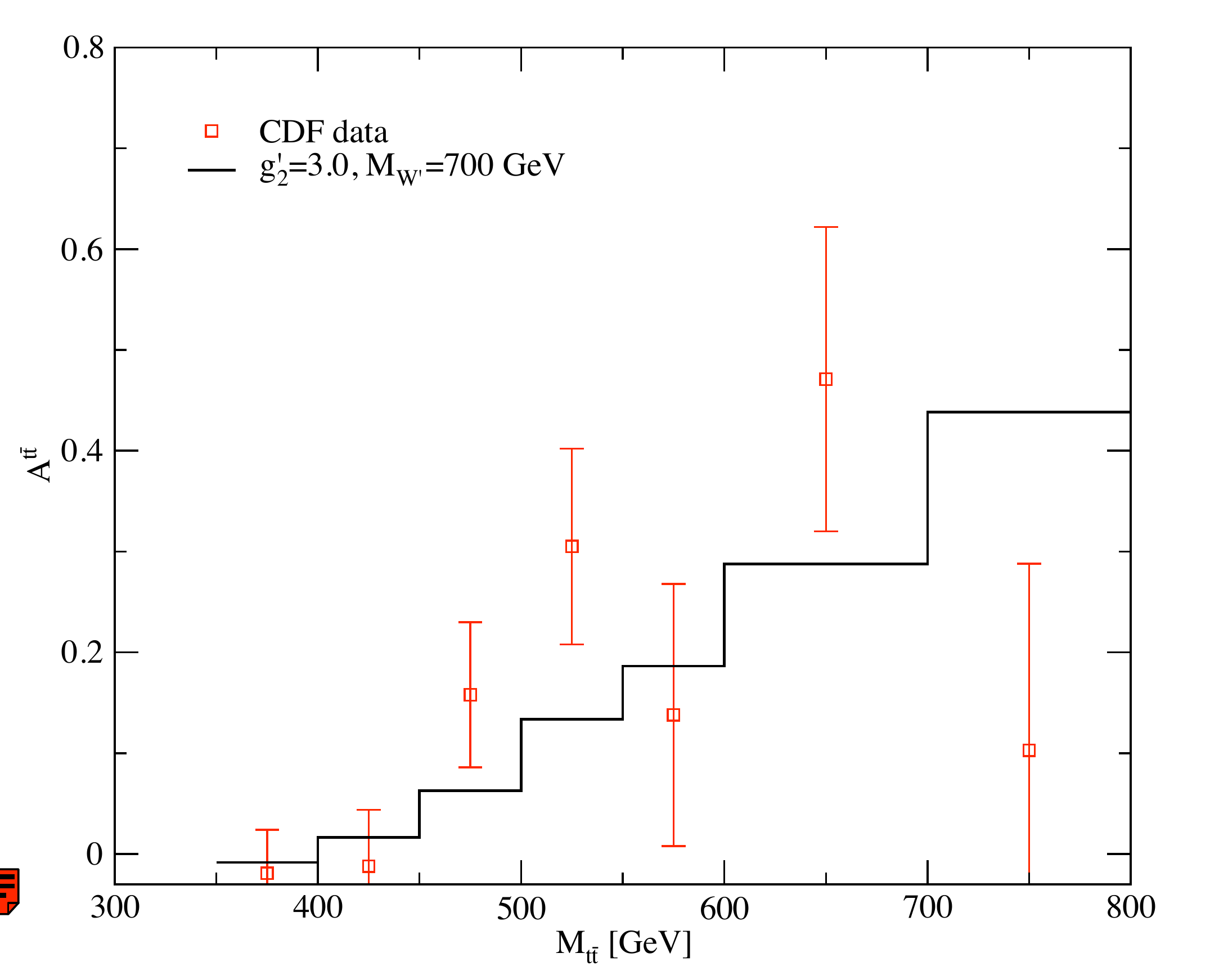}
\label{fig:mtt_afb}
}
\caption{Comparison of CDF data vs. ALRM predictions in (a) $\Delta y$ distribution and (b) $A_{FB}$ in the $p\bar p$ CM frame vs. $M_{t\bar t}$ for the model benchmark point of $g'_2=3,~M_{W'}=700$ GeV, and $M_{Z'}=1$ TeV. The points with the uncertainties are the CDF measurements and the solid histograms are the ALRM predictions including the SM QCD contribution.}
\end{figure}

\begin{figure}[htbp]
\begin{center}
\includegraphics[scale=0.5]{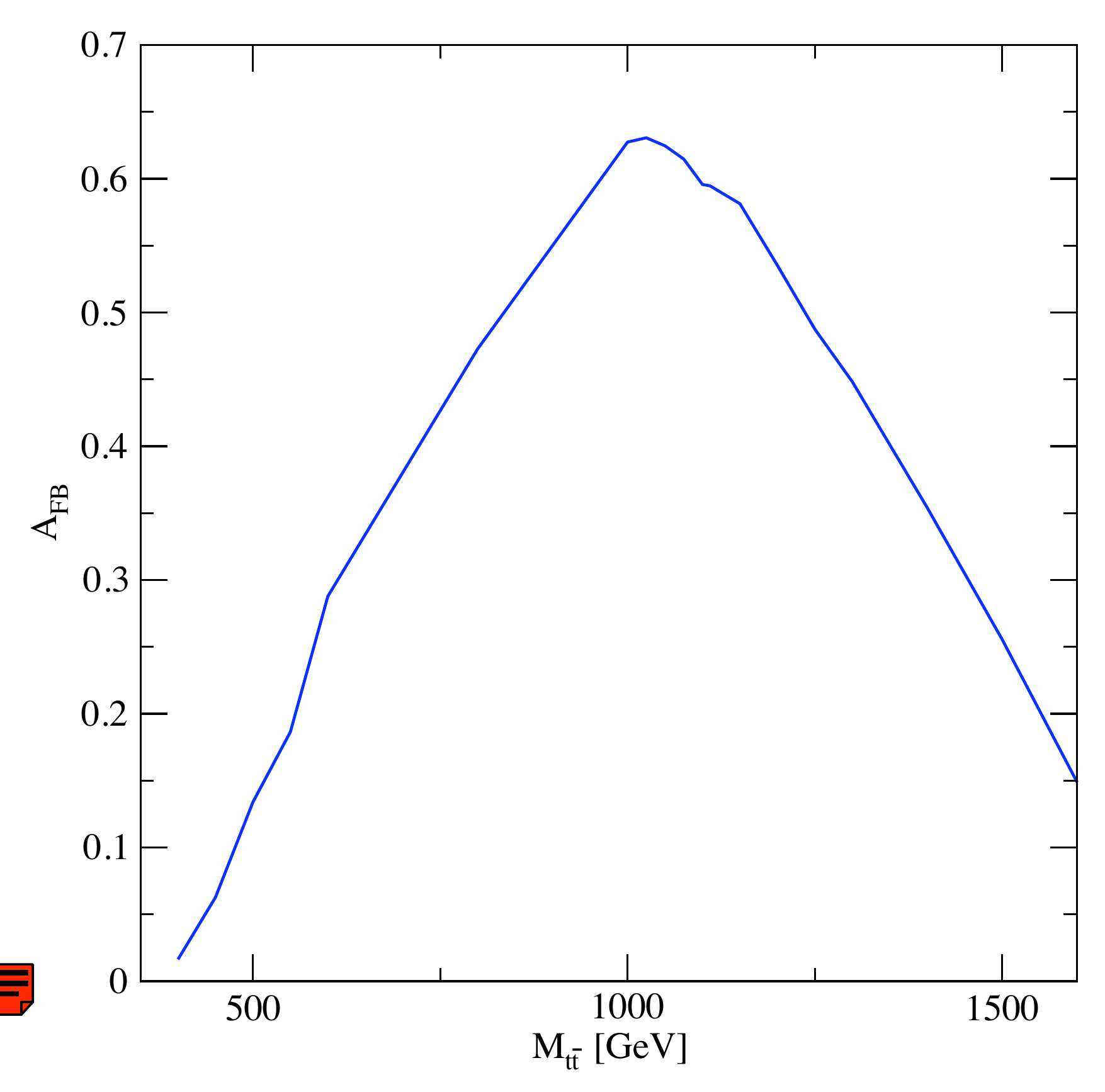}
\caption{The asymmetry in the $t\bar t$ CM frame vs. the invariant $t\bar t$ mass in the ALRM for $g'_2=3.0,~M_{W'}=700$ GeV, and $M_{Z'}=1$ TeV. The decrease in $A_{FB}$ above 1 TeV is due to the dominance of the $Z'$ contribution to the cross-section.}
\label{fig:mtt_best_high}
\end{center}
\end{figure}

\begin{figure}[htbp]
\begin{center}
\includegraphics[scale=1.2]{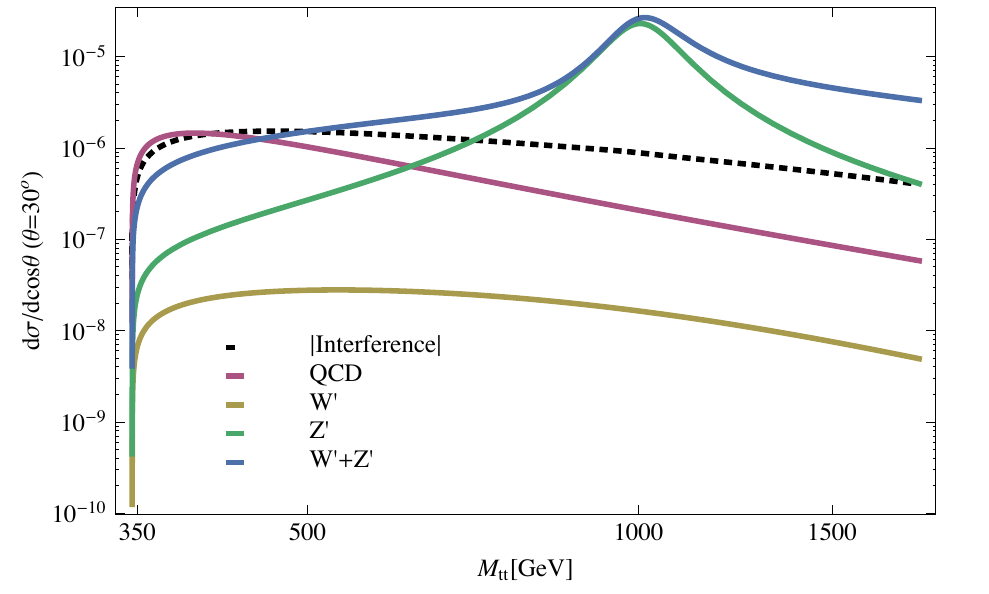}
\caption{Individual contributions to $d_R\bar d_R\to t\bar t$ at a CM scattering angle of $30^o$. We plot the magnitude of the interference term. The $Z'$ contribution dominates the subprocess in the 1 TeV region. The parton distributions emphasize the low $\hat s$ region.}
\label{fig:ddtt_sq}
\end{center}
\end{figure}

We calculate the total cross-section $\sigma(t\bar t)$, total $A_{FB}$, as well as $A_{FB}$ in $|\Delta y|<1,~|\Delta y|>1,~M_{t\bar t}<450$ GeV and $450<M_{t\bar t}<800$ GeV for various values of $g'_2$ and $M_{W'}$. We compare our results to the latest experimental results from the CDF collaboration \cite{CDFAfbNew} in Table \ref{table:afb} and in Figs.\ref{fig:mtt}\ref{fig:deltay},\ref{fig:mtt_afb}. We find that the values of $g'_2=3$, $M_{W'}=700$, and $M_{Z'}=1$ TeV provides an overall description of the data (see Table \ref{table:afbchi2}).  Alternate values of $M_{W'}$ and $g'_2$ allow for a closer match in the high invariant mass bin at the expense of consistency with the measured cross-section. We do not attempt a global fit to the data since there are correlations amongst the experimental distributions. The exact diagonalization of the weak boson mass matrices and radiative corrections to $M_{W'}$ may help alleviate this tension. We examine the dependence of $A_{FB}$ on $M_{t\bar t}$ in Fig.\ref{fig:mtt_best_high}. The shape of the curve is dominated by the $Z'$ contribution to the $d_R\bar d_R\to t\bar t$ subprocess, as shown in Fig. \ref{fig:ddtt_sq}. The structure function product $[xd(x)]^2$ has a maximum at $M_{t\bar t}=300$ GeV.

%\begin{figure}[htbp]
%\begin{center}
%\includegraphics[scale=1.2]{gp_sigma_triplet}
%\caption{$\sigma(p\bar p \to t\bar t)$ [pb] vs. $g'_2$ using $M_{Z'}=900$ GeV and the mass relation described in Eq. \ref{mwpmzprelationTriplet}. Shown are the $1\sigma$ (orange) and $2\sigma$ (yellow) bounds from experiment. We have included $K=1.3$ to account for QCD corrections.}
%\label{fig:gp_sigma_triplet}
%\end{center}
%\end{figure}

\begin{table}[htdp]
\caption{ALRM predictions for the $t\bar t$ total cross-section $\sigma(t\bar t)$, the forward-backward asymmetry in the $p\bar p$ CM frame ($A_{FB}$), and the cross-sections for the specified ranges of rapidity differences $\Delta y$ and $M_{t\bar t}$ invariant mass ranges. $M_{Z'}$ is determined by Eq.\ref{mwpmzprelationTriplet}. A QCD correction factor $K=1.3$ is included in the cross-section calculation. The ALRM asymmetry numbers are the new physics contributions only and do not include the SM QCD contribution, so they should be compared with the final row in the table. The SM values are based on the MCFM study of Ref.\cite{MCFM}. The last row is the New Physics (NP) contribution inferred from the differences of data and SM entries.}
\begin{center}
\begin{tabular}{|c|c|c|c|c|c|c|c|}
\hline
&&&&$A_{FB}$&$A_{FB}$&$A_{FB}$&$A_{FB}$\\
$g'_2$&$M_{W'}$ [GeV]&$\sigma(t\bar t)$ [pb]&$A_{FB}$&$M_{t\bar t}<450$ GeV&$450<M_{t\bar t}<800$ GeV&$|\Delta y|<1$&$|\Delta y|>1$\\\hline\hline
3.0&700&8.45&0.06&-0.01&0.136&0.03&0.14\\
3.5&700&9.05&0.11&0.01&0.22&0.06&0.26\\
3.5&650&9.8&0.16&0.03&0.26&0.06&0.36\\
3.0&550&10.4&0.22&0.04&0.33&0.09&0.42\\
2.5&500&10.5&0.19&0.003&0.32&0.07&0.40\\
\hline\hline
Data \cite{CDFAfbNew}\cite{Aaltonen:2010ic}&&\scriptsize{$7.70\pm0.52$}&\scriptsize{$0.158\pm 0.074$}&\scriptsize{$-0.116\pm0.153$}&\scriptsize{$0.475\pm0.122$}&\scriptsize{$0.026\pm 0.118$}&\scriptsize{$0.611\pm 0.256$}\\
SM&&\scriptsize{$7.45^{+0.72}_{-0.63}$}&\scriptsize{$0.058\pm0.009$}&\scriptsize{$0.04\pm0.006$}&\scriptsize{$0.088\pm0.0013$}&\scriptsize{$0.039\pm0.006$}&\scriptsize{$0.123\pm0.018$}\\
NP&&--&\scriptsize{$0.100\pm0.074$}&\scriptsize{$-0.156\pm0.147$}&\scriptsize{$0.387\pm0.121$}&\scriptsize{$0.387\pm0.112$}&\scriptsize{$0.488\pm0.257$}\\
\hline
\end{tabular}
\end{center}
\label{table:afb}
\end{table}%

\begin{table}[htdp]
\caption{$\chi^2/d.o.f.$ values for various $g'_2$ and $M_{W'}$ mass values using $A_{FB}$ in the 7 $M_{t\bar t}$ bins and the total cross-section $\sigma(t\bar t)$. We have included a $K-$factor of 1.3 for $\sigma(t\bar t)$.}
\begin{center}
\begin{tabular}{|c|c||c|c|}
\hline
$g'_2$&$M_{W'}$ [GeV]&$\chi^2~(A_{FB}^{t\bar t})/$bin&$\chi^2~(\sigma(t\bar t))$\\\hline\hline
3.0&700&1.8&2.0\\
3.5&700&1.2&6.8\\
3.5&650&1.4&15.2\\
3.0&550&1.3&27.8\\
%2.5&5000&9.7&28.1\\
\hline
\end{tabular}
\end{center}
\label{table:afbchi2}
\end{table}%

%\begin{table}[htdp]
%\caption{$\chi^2$ values for the cross-section $\sigma(t\bar t)$ and $A_{FB}$ for selected points in the ALRM with $M_{Z'}=900$ GeV.}
%\begin{center}
%\begin{tabular}{|cc|cc|cc|cc|}
%\hline
%&&$\chi^2$&&$\chi^2$&&$\chi^2$&\\
%$M_{W'}$ [GeV]&$g'_2$&$\sigma(t\bar t)$ &$A_{FB}$&$|\Delta y|<1$&$|\Delta y|>1$&$M_{t\bar t}<450$ GeV&$M_{t\bar t}>450$ GeV\\ \hline\hline
%630.3&2.5&0.5&3.3&0.1&4.1&0.4&11.2\\
%632.2& 3.0 &0.2& 0.2& 0.02&1.9&0.7&5.1\\ 
%632.6&3.2&8.2&0.1&0.2&1.9&1.1&4.5\\
%633.2&3.5&23.1&0.2&0.3&0.8&1.1&2.0\\
%\hline
%\end{tabular}
%\end{center}
%\label{table:afbchi2}
%\end{table}%

%\begin{figure}[htbp]
%\begin{center}
%\subfigure[$\Delta y$ distributions in data vs. ALRM prediction]{
%\includegraphics[scale=0.3]{deltay}
%}
%\subfigure[Parton level asymmetries at small and large $\Delta y$. The shaded bands represent the total uncertainty in each bin and the thick line is the ALRM prediction for $M_{W'}=632.2$ GeV and $g'_2=3.0$.]{\includegraphics[scale=0.3]{rapidity_afb2}
%}
%\subfigure[The $t\bar t$ frame asymmetry in the data in bins of invariant mass $M_{t\bar t}$ compared to the prediction of ALRM.]{
%\includegraphics[scale=0.3]{mtt_afb}
%}
%\caption{Comparison of the CDF data with ALRM predictions in $\Delta y$ distribution, $A_{FB}$ in small and large $\Delta y$, and $A_{FB}$ in bins of invariant mass $M_{t\bar t}$}
%\label{fig:CDFcomparison}
%\end{center}
%\end{figure}

\section{Predictions at the LHC}

\subsection{Z' signatures}
The $Z'$ in the ALRM with $\sim1$ TeV mass and $\sim200$ GeV width provides a promising route for early LHC discovery or exclusion. The $t\bar t$ invariant mass distribution should show a prominent broad peak at the $Z'$ mass \cite{Barger:2006hm} and an excess of events compared to the SM, as illustrated in Fig. \ref{figuresX}.  The golden signal of dileptons in the usual $Z'$ searches will be difficult to utilize for the ALRM $Z'$ due to its small leptonic branching fraction.
%
%\begin{figure}
%\label{figures}
%\centering
%\subfigure[7 TeV at the LHC]{
%\includegraphics[scale= 1]{ppttbar7_triplet}
%\label{fig:ppttbar7}
%}
%\subfigure[8 TeV at the LHC]{
%\includegraphics[scale= 1]{ppttbar8_triplet}
%\label{fig:ppttbar8}
%}
%\subfigure[14 TeV at the LHC]{
%\includegraphics[scale= 1]{ppttbar14_triplet}
%\label{fig:ppttbar14}
%}
%\caption[]{$M_{t\bar t}$ distributions at the LHC for $pp\to t\bar t$ at 7,8,14 TeV}
%\end{figure}

\subsection{W' signatures}
\begin{figure}
\centering
\subfigure[~7 TeV at the LHC]{
\includegraphics[scale=0.9]{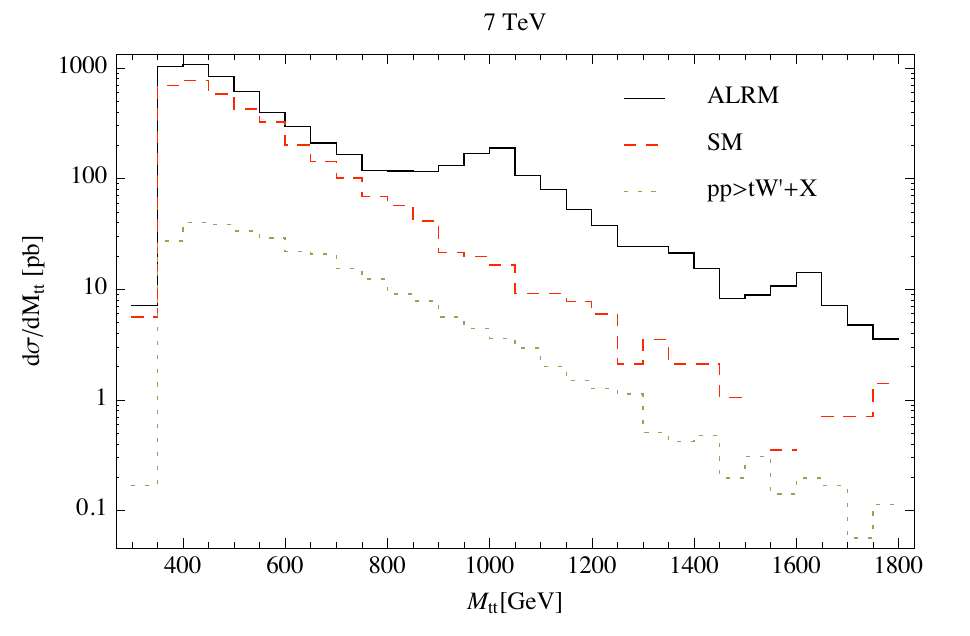}
\label{fig:ppttbarX7}
}
%\subfigure[~8 TeV at the LHC]{
%\includegraphics[scale= 0.9]{ppttbar8_triplet_best}
%\label{fig:ppttbarX8}
%}
\subfigure[~14 TeV at the LHC]{
\includegraphics[scale= 0.9]{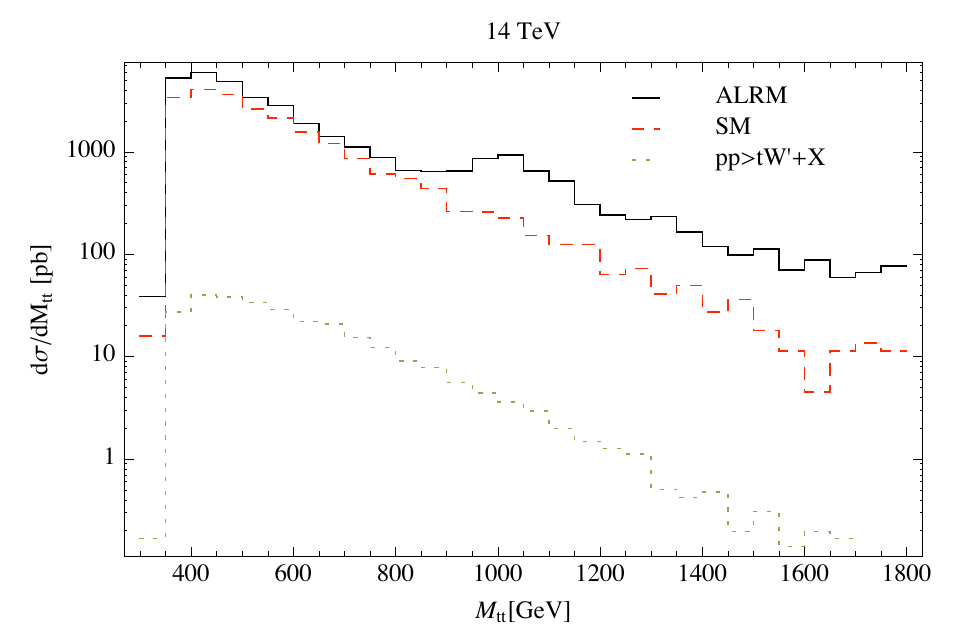}
\label{fig:ppttbarX14}
}
\caption[]{$M_{t\bar t}$ distributions at the LHC for $pp\to t\bar t+X$ at 7 and 14 TeV. There is a $Z'$ resonance peak at $M_{Z'}\sim1$ TeV as well as an excess in higher mass bins for the ALRM as compared to the SM.}
\label{figuresX}
\end{figure}

\begin{figure}[htbp]
\begin{center}
\includegraphics[scale=1]{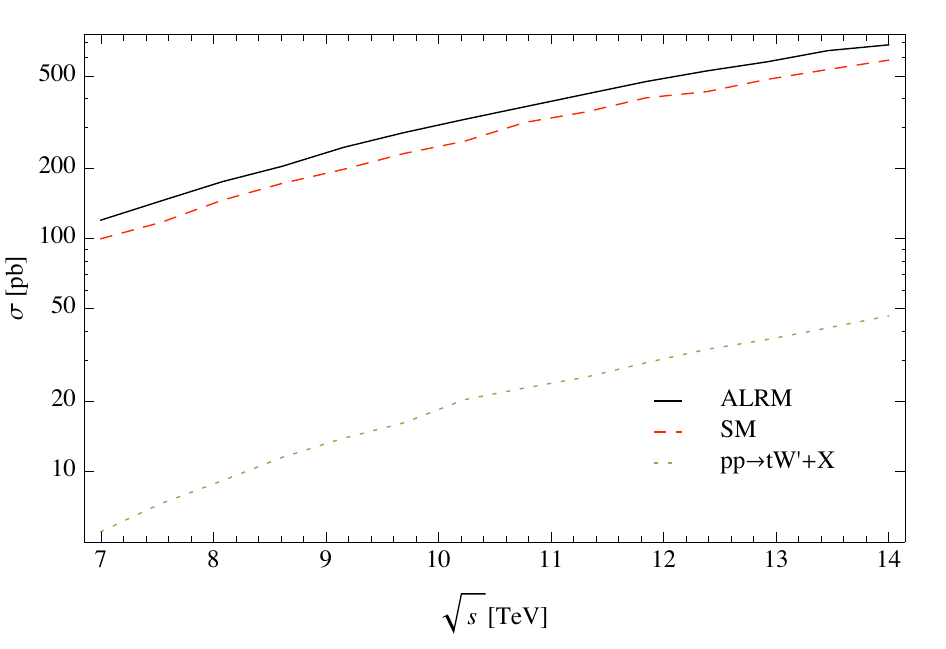}
\caption{$\sigma(pp\to t\bar t+X)$ [pb] vs. $\sqrt{s}$ [TeV] for the SM and ALRM at the LHC. Also plotted is the new physics cross-section for $pp\to tW'+X$.}
\label{fig:ttbar_energy}
\end{center}
\end{figure}

The detection of the $W'$ boson with decays to $t\bar d, \bar td$ will provide the definitive evidence for the extra $SU'(2)$ symmetry.  The subprocess for $W'$ production is $d +g \to t + W'$. The cross sections for $W'$ production in the ALRM are shown in Fig.\ref{figuresX}.  The energy dependence of the $tW'$ process is shown in Fig. \ref{fig:ttbar_energy}.  This new physics contribution to inclusive $t\bar t$ production is about $5\%$ of the SM cross section.

The search for a $W'$ at the LHC has been discussed in Refs. \cite{Barger:2010mw}\cite{Schmaltz:2010xr},\cite{Gopalakrishna:2010xm}.  The strategy for the determination of the chiral couplings of a generic $W'$ that is produced as an $s-$channel resonance and decays to $t\bar b$ was demonstrated in Ref.\cite{Gopalakrishna:2010xm}, but this process does not exist for the $W'$ in the ALRM.  A test of the right-handed nature of the $W'$ of the ALRM must be made in the process $pp\to tW'\to t(\bar td)$ which is considerably more difficult because of the ambiguity in reconstruction with two tops in the final state.

\subsection{Triplet Higgs}
We have assumed a complex triplet Higgs field $\phi'(T' =1, Y'/2 = 1)$ in order to generate a large mass splitting of the $W'$ and $Z'$ bosons of the ALRM. Thereby, the $W'$ can be sufficiently light to explain the asymmetry data, while the $Z'$ can be sufficiently heavy to escape the $Z'$ mass bounds in the $t\bar t$ channel at the Tevatron.  The $\phi'$ has no SM couplings.  The physical states of the Higgs triplet are $(\chi^{++}, \chi^+, \chi^0)$.  The detection of Higgs triplet states in hadronic collisions has been discussed in the literature  (see Ref e.g. \cite{Godfrey:2010qb}), but not for a model like the ALRM where the triplet Higgs states only couple to pairs of W' or Z' bosons.  The tell-tale signature for a complex Higgs triplet is its doubly charged members $\chi^{++}$ and $\chi^{--}$, which will decay to $W'W'$.  However, their production will require the $W'W'$ fusion process, which will be suppressed by the high $W'$ mass and by the presence of two top quarks from the $d\to tW'$ transitions in the primary fusion process.  SLHC energies are likely to be necessary to study this channel.

\section{Conclusions}
The measurement by the CDF collaboration of the forward-backward asymmetry ($A_{FB}$) of $t\bar t$ pairs exceeds the SM expectation and could be smoking gun evidence of new physics. A number of theoretical scenarios have been considered to explain the anomalous asymmetry.  Our proposal is a new gauge symmetry $U'(1)\times SU'(2)\times SU(2)$, which is broken to the SM by the neutral member of a complex Higgs triplet field.  We show that the predicted $A_{FB}$ in this model tracks the data in a $A_{FB}$ vs. $M_{t\bar t}$ plot, with a $A_{FB} $ that grows with $M_{t\bar t}$ over the CDF range.  The characteristics of the CDF measurements of
$A_{FB}$, the $M_{t\bar t}$ distribution, the rapidty difference, and the total $t\bar t$ cross section can be reproduced.  The preferred masses are 
$M_{W'} = 700$ GeV and $M_{Z'} = 1$ TeV.  The $W'$ has a right handed $(t,d)$ coupling and the $Z'$ is coupled mainly to $t\bar t$ and $d\bar d$.  The discovery of the $Z'$ with large $t\bar t$ and $d\bar d$ couplings at the LHC should be possible with early LHC data and will definitively test our proposed $SU'(2)$ gauge symmetry. A recent paper \cite{sheltonzurek} has investigated the implications of flavor violation in the ALRM with maximal mixing of $(t,d)_R$ and $(u,b)_R$ doublets. They find consistency with the constraints of flavor physics

\section{Acknowledgments}
The authors would like to thank Martin Schmaltz and Christian Spethmann for stimulating correspondence. V.B. thanks the University of Hawaii for hospitality. W.-Y. K. thanks BNL for hospitality. This work was supported in part by the U.S. Department of Energy under grants Nos. DE-FG02-95ER40896 and DE-FG02-84ER40173.  C.-T. Yu is supported by the National Science Foundation.

\end{document}